\documentclass[12pt]{iopart}
\newcommand{\be}{\begin{eqnarray}}
\newcommand{\ee}{\end{eqnarray}}
\newcommand{\del}[2]{\frac{\partial #1}{\partial #2}}
\newcommand{\vect}[1]{\mbox{\boldmath$#1$}}
\begin{document}

\letter{On the Mechanical Properties of Matter Waves}
\author{Nicholas K Whitlock$^{\dag}$, Stephen M Barnett and John Jeffers}
\address{Department of Physics, University of Strathclyde, Glasgow, G4 0NG, UK}
\ead{$^{\dag}$\mailto{whitlock@phys.strath.ac.uk}}
\pacno{03.75.-b}

\begin{abstract}
Electromagnetic waves and fluids have locally conserved mechanical
properties associated with them and we may expect these to exist for
matter waves. We present a semiclassical description of the continuity
equations relating to these conserved properties of matter-waves and
derive a general expression for their respective fluxes.
\end{abstract}

\nosections
It has long been known that electromagnetic waves have associated
mechanical properties. Maxwell described the radiation pressure
exerted by light \cite{maxwell:treatise:v2} and Poynting identified the
momentum carried by light \cite{poynting:1884:1} and that angular
momentum was associated with circular polarisation
\cite{poynting:1909:1}. The energy, momentum and angular momentum of
electromagnetic waves are particularly important as these quantities
are all conserved. Associated with the density of each of these
quantities is a flux density (or current density) which is related to
the density via continuity equations \cite{barnett:2002:1}. 

The creation of the first Bose-Einstein condensates (BECs) in
dilute atomic gases in 1995 \cite{anderson:1995:1, davis:1995:1,
  bradley:1995:1} brought about the possibility of creating coherent
matter waves. A BEC in a trap is analogous to an electromagnetic
field in a laser cavity, in that macroscopic occupation of a single
quantum state occurs. Just as a laser cavity is a source of
coherent light, a BEC can be a source of coherent matter waves. To
continue this analogy we assert that matter waves will have associated
mechanical properties and present a general theory of the local
mechanical properties of matter waves. We give, as examples, the
densities and flux densities of the matter wave particle number
and energy. The same theory applies to their momentum and angular
momentum.

The local conservation of particles is a well established concept in
fluid mechanics. It states that in any volume element of a
fluid, no matter how small, the rate of change of particles
in an element of fluid is equal to the rate at which the particles
flow into the surface of the element. This principle is made
quantitative in the following equation \cite{landlif:v6};
\be
\del{}{t}\int_V\rho\rmd V = -\oint_S\rho\vect{v}\cdot\rmd\vect{S},
\ee
where $\rho$ is the particle density, $\vect{v}$ is the velocity of
the fluid  and $\rmd\vect{S}$ is a surface element with direction normal
to the surface enclosing the volume $V$. If we define the particle
current density (or {\it flux density}) $\vect{J} = \rho\vect{v}$,
then we can easily obtain the continuity equation for particles in a
fluid
\be
\del{\rho}{t} + \nabla \cdot \vect{J} = 0. \label{eq:fluidpart}
\ee
Similar continuity equations hold for other conserved quantities
\cite{landlif:v6} such as the energy density
\be
\del{w}{t} + \nabla\cdot\left[\rho\vect{v}\left(\frac{v^2}{2} +
\epsilon + \frac{P}{\rho}\right)\right] = 0, \label{eq:fluiden}
\ee
where $\epsilon$ is the internal energy per unit mass and $P$ is the
pressure of the fluid, and the momentum density
\be
\del{p_i}{t} + \del{}{x_j}\left(P\delta_{ij} + \rho v_iv_j\right) = 0.
\label{eq:fluidmtm}
\ee
Here the summation convention has been introduced, so the indices ($i$
and $j$) take the values 1, 2, 3 and a repeated index is understood to
be summed. This convention will be assumed for the remainder of the
paper.

In a case in which a globally conserved quantity does not obey the
continuity equation of the form (\ref{eq:fluidpart}), there will be a
source of that quantity. An example of this is a fluid under the
influence of an external force, which is a source of momentum. In such
a system it is necessary to modify the equation of continuity to allow
for the source term. We would expect that the rate of change of
momentum in an element of volume would be equal to the rate at which
momentum flows into the surface plus the rate at which momentum is
created in the element. This reasoning leads us to a modified
continuity equation of the form
\be
\del{p_i}{t} + \del{}{x_j}\Pi_{ij} = S_i,
\ee
where $\Pi_{ij}$ is the $\{i,j\}$ component of the momentum flux
density tensor and $S_i$ is the source density for the $i$ component of
momentum.

We now turn our attention to the semiclassical description of
matter waves. Consider an observable $A$ having a corresponding
Hermitian operator $\hat{A}$. The density
$\hat{\mathcal{A}}(\vect{R})$ of $A$ at a point $\vect{R}$ will obey
the generalised local continuity equation
\be
\del{}{t}\hat{\mathcal{A}}(\vect{R}) +
\del{}{X_i}\hat{\mathcal{T}}^{(A)}_i(\vect{R}) =
\hat{\mathcal{S}}^{(A)}(\vect{R}), \label{eq:cons}
\ee
where $X_i$ are the components of $\vect{R}$,
$\hat{\mathcal{T}}_i^{(A)}$ is the flux density of the observable and
$\hat{\mathcal{S}}^{(A)}$ is the source density term. We define the
Hermitian operator associated with the density of $A$ at $\vect{R}$ to
be
\be
\hat{\mathcal{A}}(\vect{R}) =
\frac{1}{2}\left\{\hat{A},\delta(\hat{\vect{r}}-\vect{R})\right\},
\label{eq:a}
\ee
where $\{,\}$ denotes the anticommutator and
$\delta(\hat{\vect{r}}-\vect{R})$ is the Dirac delta function. From the
Heisenberg equation of motion we can show that the density will evolve
according to
\be
\del{}{t}\hat{\mathcal{A}}(\vect{R}) &=& \frac{\rmi}{2\hbar}
\left(\left\{\left[H,A\right],\delta(\hat{\vect{r}}-\vect{R})\right\} +
\left\{\hat{A},\left[H,\delta(\hat{\vect{r}}-\vect{R})\right]
\right\}\right) \label{eq:timeder}
\ee
We see that the first term on the right hand side will vanish if $A$
is conserved. It is therefore reasonable to associate this term with
the source term in (\ref{eq:cons}) and the last term with the flux
density. In order to proceed in the calculation we use the
standard Hamiltonian for a particle in a potential $V(\vect{r})$
\be
\hat{H} = -\frac{\hbar^2}{2m}\nabla_i\nabla_i + V(\vect{r}), \label{eq:ham}
\ee
where $\nabla_i\equiv\del{}{x_i}$. On evaluating the last term in
(\ref{eq:timeder}) we find that
\be
\fl\frac{1}{2}\del{}{t}\left\{\hat{A},\delta(\hat{\vect{r}}-\vect{R})\right\} +
\frac{1}{4}\nabla_{R,i}
\left\{\hat{A},\left\{\frac{\hat{p_i}}{m},\delta(\hat{\vect{r}}-\vect{R})
\right\}\right\} =
\frac{1}{2}\left\{\del{}{t}\hat{A},\delta(\hat{\vect{r}}-\vect{R})\right\},
\label{eq:anticommcont}
\ee
which takes the form of a conservation equation (\ref{eq:cons}). This
allows us to identify the flux density operator
\be
\hat{\mathcal{T}}_i^{(A)}(\vect{R}) = \frac{1}{4}
\left\{\hat{A},\left\{\frac{\hat{p_i}}{m},\delta(\hat{\vect{r}}-\vect{R})
\right\}\right\}. \label{eq:ambig}
\ee
We can associate the operator $\left\{\hat{p_i}/m,\delta(\hat{\vect{r}}-\vect{R})
\right\}$ with the velocity density and hence see that
$\hat{\mathcal{T}}_i^{(A)}(\vect{R})$ is a suitably symmetrised
product of the velocity density and the observable. This operator
is the flow of the density of the observable, that is its flux
density. The forms of the flux density and source density for the
observable $A$ will come from the expectation value of
(\ref{eq:anticommcont}). In the Heisenberg picture, the matter wave
will be described by a wavefunction $\psi(\vect{r})$. The expectation
value of the conservation equation for this state is
\be
\del{}{t}\mathcal{A}(\vect{R}) +
\nabla_{R,i}\mathcal{T}_i^{(A)}(\vect{R}) =
\mathcal{S}^{(A)}(\vect{R}), \label{eq:avcons}
\ee
which is the continuity equation for the expectation values of the
relevant operators. Here we have introduced the notation
$\mathcal{A}(\vect{R}) \equiv
\langle\hat{\mathcal{A}}(\vect{R})\rangle$. The flux density
expectation value is
\be
\fl\mathcal{T}_i^{(A)}(\vect{R}) &=&
\int\rmd^3r\psi^*(\vect{r})\hat{\mathcal{T}}_i^{(A)}(\vect{R})
\psi(\vect{r})\nonumber\\
\fl&=&\frac{\rmi\hbar}{4m}\left.\left[\psi\nabla_i(\hat{A}\psi)^* -
(\hat{A}\psi)^*\nabla_i\psi + \hat{A}\psi(\nabla_i\psi^*) -
\psi^*\nabla_i(\hat{A}\psi)\right]\right|_{\vect{r}=\vect{R}}
\label{eq:fluxdens}
\ee
and the source density expectation value is
\be
\mathcal{S}^{(A)}(\vect{R}) = \frac{\rmi}{2\hbar}
\left.\left\{-\left(\left[\hat{H},\hat{A}\right]\psi
\right)^*\psi + \psi^* \left(\left[\hat{H},\hat{A}\right]\psi\right)
\right\}\right|_{\vect{r}=\vect{R}}. \label{eq:sourcedens}
\ee
As an example we consider two quantities which obey continuity
equations, specifically the particle number and energy. Following Landau
\cite{landau:1941:1}, the operator corresponding to particle density
in a semiclassical theory is represented by the Dirac delta function,
so in (\ref{eq:a}) we put the operator $\hat{A}$ equal to the identity
operator $\hat{1}$. From (\ref{eq:fluxdens}) the flux density of
particles at position $\vect{r}$ can be calculated as
\be
\mathcal{T}^{(\rho)}_i(\vect{R}) =
\frac{\hbar}{m}\mathrm{Im}\left.\left\{\psi^*\nabla_i\psi
\right\}\right|_{\vect{r}=\vect{R}},
\ee
where Im indicates the imaginary part. This is the form of the
well-known quantum mechanical probability flux $J_i$
\cite{bransden:book}. The source term
vanishes as the identity operator commutes with the Hamiltonian. This
is a reasonable result as there is no particle source present in the
system. Energy in quantum mechanics is represented by the Hamiltonian
operator, so we can calculate the energy flux
density from (\ref{eq:fluxdens}). If we use the Hamiltonian
(\ref{eq:ham}) we obtain the energy flux density at position
$\vect{R}$, which is
\be
\mathcal{T}^{(E)}_i(\vect{R}) =
\frac{\hbar^3}{4m^2}\mathrm{Im}\left.\left\{\psi\nabla^2\nabla_i\psi^* +
\left(\nabla_i\psi^*\right)\nabla^2\psi\right\}\right|_{\vect{r}=\vect{R}} +
V(\vect{R})\mathcal{T}^{(\rho)}_i(\vect{R}). \label{eq:quanten}
\ee
The source term vanishes as the Hamiltonian commutes with itself.

In this letter, we have derived general flux density and
source density operators for conserved quantities in matter waves
within the semiclassical description. This theory will be important
when considering the deposition of these quantities during
interactions of matter waves with other objects. An example of this is
the detection of matter waves themselves, which requires one to
consider the particle flux density \cite{whitlock:2003:1}. One problem
is that the flux density operator, expression (\ref{eq:ambig}), might
contain an arbitrary divergenceless operator in addition to the terms
present. It seems likely, as with Poynting's theorem, that this
freedom will have no physical consequences \cite{jackson:book}. Also
of interest is the form of the second-quantised version of the theory
presented here. We will address these issues elsewhere.

\ack
This work was funded in part by the Overseas Research Students Award
Scheme and the University of Strathclyde.

\end{document}